\documentclass[11pt]{article}
\usepackage{acl2017}
\usepackage{times}
\usepackage{url}
\usepackage{latexsym}
\usepackage{xspace}
\usepackage{booktabs}
\usepackage{color}
\usepackage{graphicx}
\usepackage{tabulary}
\usepackage{enumitem}

\usepackage{amsfonts}

\newenvironment{tightitemize}%
  {\begin{itemize}[topsep=0pt, partopsep=0pt] %
    \setlength{\itemsep}{0pt}%
    \setlength{\parskip}{0pt}%
    }%
  {\end{itemize}}
  \newenvironment{tightitemize1}%
  {\begin{itemize}[topsep=0pt, partopsep=0pt]%
    \setlength{\itemsep}{0pt}%
    \setlength{\parskip}{0pt}%
    }%
  {\end{itemize}}  
\usepackage{multirow}
\aclfinalcopy

\newenvironment{tightenu}%
  {\begin{enumerate}[topsep=0pt, partopsep=0pt] \footnotesize%
    \setlength{\itemsep}{0pt}%
    \setlength{\parskip}{0pt}%
    }%
  {\end{enumerate}}

  \usepackage{color}
  {\begin{enumerate}≈ \footnotesize%
    \setlength{\itemsep}{0pt}%
    \setlength{\parskip}{0pt}%
    }%
  {\end{enumerate}}

\usepackage{microtype}
\usepackage{multirow}
\usepackage{verbatim}
\usepackage{amsmath,amsthm,amssymb}
\usepackage{array}
\usepackage[scaled=0.86]{helvet}
\usepackage{ifthen}
\usepackage{courier}
\usepackage[linesnumbered,vlined,ruled]{algorithm2e}

\newcommand{\captionfonts}{\small}
\makeatletter  
\long\def\@makecaption#1#2{%
  \vskip\abovecaptionskip
  \sbox\@tempboxa{{\captionfonts #1: #2}}%
  \ifdim \wd\@tempboxa >\hsize
    {\captionfonts #1: #2\par}
  \else
    \hbox to\hsize{\hfil\box\@tempboxa\hfil}%
  \fi
  \vskip\belowcaptionskip}
\makeatother   

\belowdisplayskip \abovedisplayskip

\DeclareMathOperator*{\argmin}{arg\,min}
\DeclareMathOperator*{\argmax}{arg\,max}

\title{Learning Multi-faceted Representations of Individuals from Heterogeneous Evidence using Neural Networks}

\author{Jiwei Li, Alan Ritter and Dan Jurafsky \\
Stanford University, Stanford, CA, USA \\
Ohio State University, OH, USA \\
{\tt jiweil,jurafsky@stanford.edu,ritter.1492@osu.edu} \\
}
\date{}

\begin{document}
\maketitle

\begin{abstract}
Inferring latent attributes of people online is an important social computing task, but
requires integrating the many heterogeneous sources of information available on the web.
We propose learning individual representations of people using
neural nets to integrate rich linguistic and network evidence gathered from social media.
The algorithm  is able to combine diverse cues, such as
the text a person writes, their attributes (e.g. gender, employer, education, location) and
social relations to other people.
We show that by integrating both textual and network evidence, these representations offer 
improved performance at four important tasks
in social media inference on Twitter: predicting (1) gender, (2) occupation, (3) location,
and (4) friendships for users.
Our approach scales to large datasets and
the learned representations 
can be used as general features in and
have the potential to benefit a large number of downstream tasks
including link prediction, community detection,
or probabilistic reasoning over social networks.
\end{abstract}

\section{Introduction}

The recent rise of online social media presents an unprecedented opportunity for computational social science:
user generated texts provide insight about user's attributes such as employment, education or gender.  
At the same time the social network structure sheds light on complex real-world relationships between preferences and attributes.
For instance  people sharing similar
attributes such as employment background or hobbies have a higher chance of becoming friends.
User modeling based on information presented in social networks is an important goal, both for applications such as product recommendation, 
targeted online advertising, friend recommendation and for helping social scientists and political analysts gain insights into public opinions and user behaviors. 

\begin{figure}
\centering
\includegraphics[width=3in]{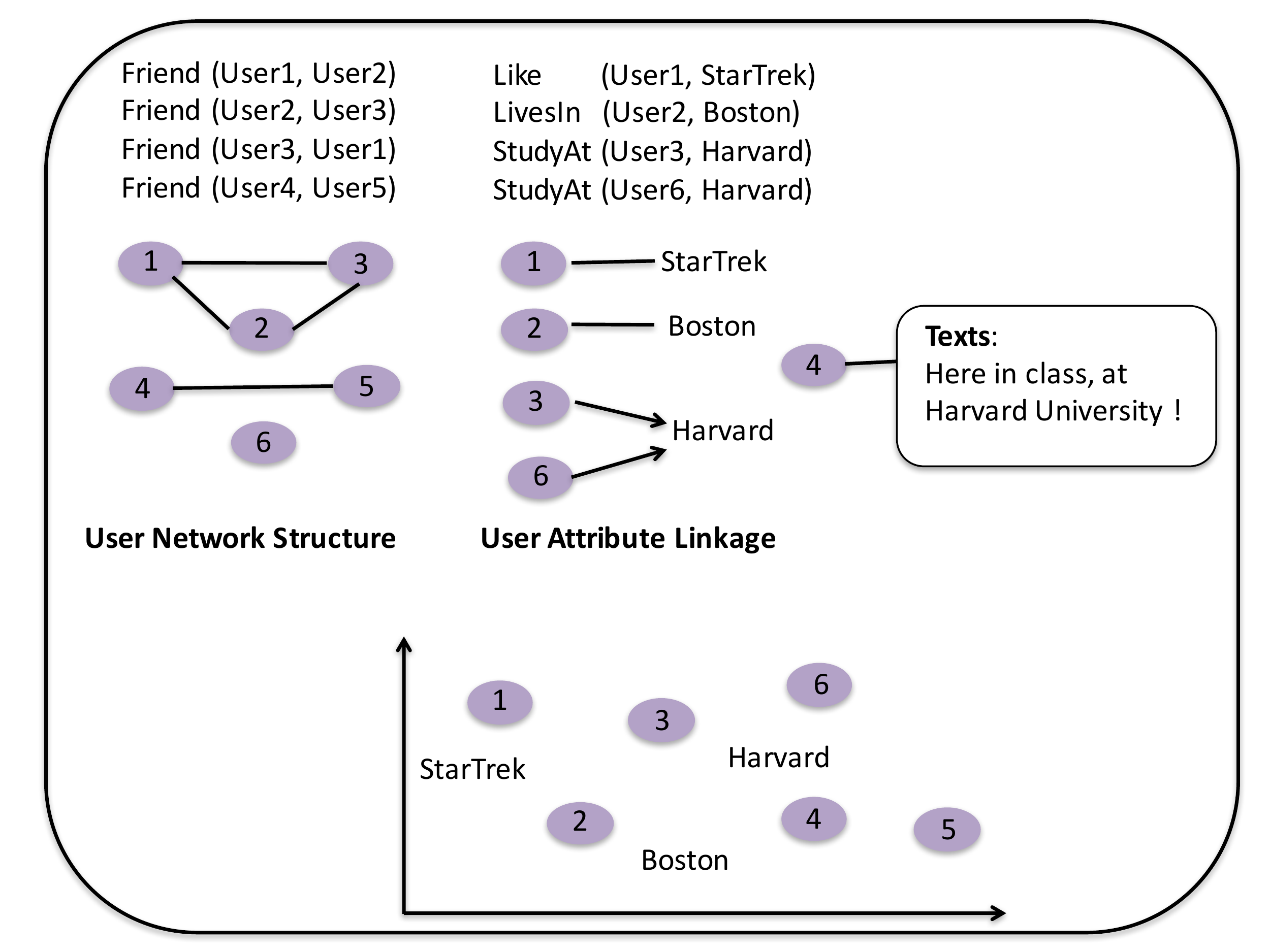}
\caption{Illustration for the proposed method that learns latent representations for users, attributes and user-generated texts based on social network information.}\label{brief}
\end{figure}

Nevertheless, much important information on social networks exists in unstructured data formats.
Important social insights are locked away, entangled within a heterogenous combination of social signals \cite{sun2009ranking} - including text, networks, attributes, relations, preferences, etc.
While recent models have attempted to link one or two aspects of the evidence, how to develop a scalable framework that
 incorporates massive, diverse social signals including user generated texts, tens of thousands of user attributes and network structure in an integrated way, 
 remains an open problem.

In this paper, we propose a general deep learning framework for jointly analyzing user networks, generated context and attributes.
We map users, attributes, and user-generated content to latent vector representations, which are learned in a scalable way from social network data.
Figure \ref{brief} gives a brief overview of the mechanism of the proposed model: 
users are represented by similar vectors if they are friends, share similar attributes or write similar content. Attributes are similarly clustered if 
associated with similar users.\footnote{This core idea is similar to collaborative filtering \cite{kautz1997referral}.} 
In summary, we incorporate diverse social signals in a  unified framework, allowing user embeddings to be jointly optimized 
using neural networks trained on vast quantities of rich social and linguistic context.

Based on these learned representations, our approach provides with a general 
paradigm 
 on a wide range of predictive tasks concerning individual users
as well as group behavior: user attribute inference (e.g., the city the user lives in), 
personal interest prediction (e.g, whether a user will like a particular movie),
and probabilistic logical reasoning over the social network graph. For example, our models infer that:
 \begin{tightitemize}
\item Men in California are 6.8 times more likely to take an engineering occupation than women in California. 
 \item  Users who work in the IT industry\footnote{This information comes from the Standard Occupational Classification (SOC), as will be described later.} 
are 2.5 times more likely to like iPhones than users working in Legal Occupations.
 \end{tightitemize}
Our methods also have the potential to seamlessly integrate rich textual context into many social network analysis tasks including: 
link prediction, community detection, and so on, and the learned user representations can be used as important input features for  downstream machine learning models, just as how word embeddings are used in the field of natural language processing. 
The major contributions of this paper can be summarized as follows:
\begin{tightitemize}
\item We propose new ways for integrating heterogeneous cues  about people's
relations or attributes into a single latent representation.
\item We present  inference algorithms for
solving social media inference tasks related to both individual and group behaviors based on the learned user representations.
\item Cues that we take advantage of may be noisy and sometimes absent but by combining them via global inference, we can learn latent facts about people.
\end{tightitemize}

We evaluate the proposed model on four diverse tasks: friend-relation prediction, gender identification, occupation identification and user geolocation prediction. 
Experimental results demonstrate improved predictions by our model by incorporating diverse evidence from many sources of social signals.

\section{Social Representation Learning}
Our goal is to learn latent representations
from the 
 following three types of online information: (1) user-generated texts (2) friend networks (3) relations and attributes.

\subsection{Modeling Text}
User generated texts reflect a user's interests, backgrounds, personalities, etc.
We thus propose learning user representations based on the text a user generates. 

We represent each user $v$ by a $K$-dimensional vector $e_v$. 
Suppose that $S$ denotes a sequence of tokens $S=\{w_1, w_2, ..., w_{N_S}\}$ generated by the current user $v$.
Each word $w\in S$ is associated with a $K$-dimensional vector $e_w$.
Let $C(w)$ denote the list of neighboring words for token $w$. 
$w$
is generated based on 
not only 
 a general language model shared across all users (namely, a model 
 that predicts $w$ given 
  $C(w)$), but the 
 representation $e_v$ of the
 current user :
 \begin{equation}
\begin{aligned}
&P(w|C(w), v)= p(w| e_C)\\
&e_C=\frac{1}{|C_w+1|}[\sum_{w'\in C(w)}e_{w'}+e_v]
\end{aligned}
\label{equ7}
\end{equation}
From Eq.\ref{equ7}, we are predicting the current word given the combination of  its neighbors' embeddings and the current user embedding. This is akin to the CBOW model \cite{mikolov2013efficient} with the only difference  that the user embedding is added into contexts.  
Such an idea also
resembles 
 the {\it paragraph vector} model \cite{le2014distributed} and the multimodal language model \cite{kiros2014multimodal} .

We use negative sampling, in which we randomly generate negative words $w^*$. 
Let $L_w$ denote a binary variable indicating whether the current word $w$ is generated by the current user. The loss function using negative sampling is  given by:
\begin{equation*}
\text{Loss(text)}=\log p(L_w=1|v)+\sum_{w^*}\log p(L_w=0|v)
\end{equation*}

Word prediction errors are backpropogated to user embeddings, 
pushing the representations of users who generate similar texts to be similar. 
\subsection{Modeling User Networks}
\label{graph_objective}
By the {\it homophily} effect, individuals who are friends on social networks tend to share common characteristics.  We therefore to encourage users who are friends have similar representations. 

We propose using a strategy similar to skip-gram models, in which we consider users who are friends on social networks analogous to neighboring words, 
the representations of which we wis to be similar.
On the other hand, we want embeddings of individuals who are not friends to be distant, just as words that do not co-appear in context.
A similar idea of transforming social graph to vector space embeddings has been explored in the recent {\it deepwalk} model 
\cite{perozzi2014deepwalk} 
and {\it Line} \cite{tang2015line}.

Suppose we have two users $v$ and $v'$. The probability that the two users are and are not friends are respectively given by:
\begin{equation}
\begin{aligned}
&\log p(L(v,v')=1)=\log\frac{1}{1+\exp(-e_v\cdot e_{v'})}\\
&\log p(L(v,v')=0)=\log\frac{1}{1+\exp(e_v\cdot e_{v'})}
\end{aligned}
\label{eq-graph}
\end{equation}
From Eq \ref{eq-graph}, we can see that the model favors the cases where the dot product of friends' embeddings is large, equivalent to 
 their embeddings being similar.
Again, we use negative sampling for optimization. For two users $v$ and $v'$ who are friends, we  sample $N$ random users $v*$, and we assume friendship does not hold between them.  The objective function is therefore given by:
\begin{equation*}
\begin{aligned}
&\text{Loss(graph)} \\
&=\log p(L(v,v')=1)+\sum_{v*}\log p(L(v,v^{*})=0)
\end{aligned}
\end{equation*}

\subsubsection{Modeling Relations and Attributes}
Intuitively, users who share similar attributes should also have similar representations and occupy similar position in the vector space. 
Suppose that a specific relation $r$ holds between a user $v$ and an entity $m$.
We represent each user and entity by a K-dimensional vector and a relation by a $K\times K$ matrix. 
For any tuple $(r, v, m)$, we map it to a scalar within the range of [0,1], indicating the likelihood of relation $r$ holding between user $v$ and entity $m$:
\begin{equation*}
\begin{aligned}
&\log p(L(r, v, m)=1)\\
&~~~~~~~~~~~~~~=\log\frac{1}{1+\exp(-e_{v_i}^T \cdot R_i\cdot e_m)}\\
\end{aligned}
\end{equation*}
Similar scoring functions have been applied in a variety of  work for relation extraction   \cite{socher2013reasoning,chang2014typed}. 
Again we turn to negative sampling for optimization. 
The system randomly samples none-referred-to entities and maximizes the difference between the observed relation tuples and 
randomly sampled ones.
Through the model described above, users who share similar attributes will have similar representations. 
At the same time, entities that are shared by similar users would also
have similar representations and thus
 occupy close positions in the vector space.  
\subsection{Training}
The global learning objective is a linear combination of the objectives from the three categories described above. User embeddings are shared across these categories, and each part can communicate with the rest: a user who publishes content about a particular city (text modeling) can have similar embeddings to those who live in that city (relation/attribution modeling);  friends (graph modeling) of a basketball fan (relation/attribution modeling) are more likely to be basketball fans as well.  
The final objective function is  given as follows:
\begin{equation*}
\begin{aligned}
L(\Theta)=
&\text{Loss(text)}+\lambda_1\text{Loss(graph)}\\
&+\lambda_2\text{Loss(relation/attribute)}
\end{aligned}
\end{equation*}
\begin{equation*}
\begin{aligned}
&\Theta=\argmin_{\Theta'}L(\Theta')
\end{aligned}
\end{equation*}
where $\lambda_1$ and $\lambda_2$ denote weights for different constituents. 
We use 
Stochastic gradient decent \cite{zhang2004solving} to update the parameters.
\begin{equation*}
\Theta^t:=\Theta^{t-1}-\alpha\frac{\partial L}{\partial\Theta}
\label{gradient}
\end{equation*}
The system jointly learns user embeddings, word embeddings, entity embeddings and relation matrices. 
\section{Inference on Social Networks}
In this section, we describe how to take as input
 a learned user embedding  for different inference tasks on social media.    
We divide  inference tasks on social networks
into two major categories: inference for individual behaviors and group behaviors. 
The former focuses on inferring attributes of a specific user such as whether a user likes a specific entity or whether 
a specific relation holds between two users, 
while the latter focuses on inference over a group of users, e.g., what is the probability of a new yorker being a fan of Knicks.

\subsection{ User Attribute Inference}
Given user representation $e_{v_i}$, we wish to infer the label
for a specific attribute
 of a specific user. The label can be whether a user likes an entity in a binary classification task, or the state that a user lives in a multi-class classification task. 

Suppose that we want to predict an attribute label (denoted by $t_{v} \in [1,L]$) for
a user $v$.
We assume that information for this attribute is embedded in user representations  
and therefore, 
and 
build another neural model  to expose this information.
Specifically, the  model takes as input user embedding $e_v$ and outputs the attribute label  using a softmax function  as follows:
\begin{equation*}
\begin{aligned}
&h= \text{tanh}(W\cdot e_{v}) \\
&S_l=U_l\cdot h \\
&p(t_{v_i}=l)=\frac{\text{exp}(S_l)}{\sum_{l'\in [1,L]} \text{exp}(S_{l'})}
\end{aligned}
\end{equation*}
Parameters to learn include $W$ and $U$.
User representations are kept fixed during training. The model is optimized 
 using AdaGrad \cite{zeiler2012adadelta}.

\subsection{User Relation Inference}
{\it User Relation Inference} specifies whether a particular relationship holds between two users (e.g., whether the two users are friends). 
It takes 
embeddings for
 both users as inputs.
Given a user $v_i$ (associated with embedding $e_{v_i}$) and
a user $v_j$ (associated with embedding $e_{v_j}$), 
we wish to predict the index of relationship label $t(v_i,v_j)\in[1,L]$ that holds between the two users.
A 
neural network prediction model is trained 
that takes as input the embeddings of the two users.
The model considers
 the distances and angle between the two user embeddings. Similar strategies can be found in many existing works, e.g., \cite{tai2015improved}.

Non-linear composition is first applied to both user representations: 
\begin{equation*}
\begin{aligned}
&\hat{h}_{v_1}=\text{tanh}(W_a\cdot e_{v_1})\\
&\hat{h}_{v_2}=\text{tanh}(W_b\cdot e_{v_2})\\
\end{aligned}
\end{equation*}
Next the distance and angle between $\hat{h}_{v_1}$ and $\hat{h}_{v_2}$ are computed:
\begin{equation*}
\begin{aligned}
&h_{+}=\hat{h}_{v_1}\otimes \hat{h}_{v_2}\\
&h_{\times}=|\hat{h}_{v_1}-\hat{h}_{v_2}| \\
\end{aligned}
\end{equation*}
The multiplicative
measure $h_{\times}$ is the elementwise comparison of the signs of the input representations. 
Finally  a softmax function is used to decide the label:
\begin{equation*}
\begin{aligned}
&h=\text{tanh}(W^{\times}\cdot h_{\times}+W^{+}\cdot h_{+}+W_1\cdot \hat{h}_{v_1}+W_2\cdot \hat{h}_{v_2})\\
&~~~~p[t(v_i,v_j)=l]=\text{softmax}(U\cdot h)
\end{aligned}
\end{equation*}
Again, parameters involved 
 are learned using stochastic gradient decent with AdaGrad \cite{zeiler2012adadelta}. 

\subsection{Inference of Group Behavior}
We now return to the example described in Section 1, in which we wish to estimate the probability of a male located in California (Cal for short) having an engineering occupation. 
Given a list of users, their representations, and gold-standard labels, we first separately train the following neural classifiers:
\begin{itemize}
\item whether a user is a male, i.e., P(\text{gender}($e_{v_i}$)=\text{male}) 
\item whether a user lives in Cal, i.e., P(\text{LiveIn}($e_{v_i}$)=\text{Cal})  
\item whether a user takes an engineering occupation,\\ i.e., P(\text{Job}($e_{v_i}$)=\text{engineering})  
\end{itemize}
Next we estimate the general embedding (denoted by $e_G$) for the group of people that satisfy 
these premises, namely, they 
 are males and live in California at the same time. This can be transformed to the following optimization problem with $e_G$ being the parameters to learn: \footnote{Note that we assume propositions are independent, so in the example above,
being a male and living in California are independent from each other. We leave relaxing this independence assumption to future work.}
 \begin{equation}
 \begin{aligned}
&e_G=\argmax_{e} \log P(\text{gender}(e)=\text{male})\\
&~~~~~~~~~~~~~~~~~~~~~~~-\log P(\text{LiveIn}(e)=\text{Cal})
\end{aligned}
\label{opt}
\end{equation}
Eq.\ref{opt} can be thought of as  an optimization problem to find a optimal value of $e_G$.
This problem can be solved
 using SGD. 
The obtained optimal 
 $e_G$ is used to represent
 that group embedding for users who satisfy the premises (e..g, males and living in Cal). 
$e_G$  is then used as inputs to classifier 
$P(\text{Job} (e_G))=\text{engineering})$ which returns probability of taking an engineering job. 

More formally, given a list of conditions $\{a_i\}\in A$, we want to the compute the probability 
that another list of conditions (denoted by $B=\{b_j\}$) hold, in which $b_j$ can be a user being an engineer.
This probability is denoted by $p(B|A)$.
The algorithm
to compute $p(B|A)$ 
 for group behavior inference is summarized in Figure \ref{fig:generative process}.
\begin{figure}[!ht]
\small
\rule{8cm}{0.03cm}
\begin{tightitemize}
\item For $a_i\in A$, $b_j\in B$, train separate classifiers  $P(a_i|e), P(b_j|e)$ based on user representations $e$ and labeled datasets.
\item Estimate group representation $e_G$ by solving the following optimization problem using SGD:
$$e_G=\argmax_{e}\prod_{a_i\in A}P(a_i|e)$$
\item Infer the probability :
$$P(B|A)=\prod_{b_j\in B}p(b_j|e(\text{group}))$$
\end{tightitemize}
\rule{8cm}{0.03cm}
\caption{Algorithm for group behavior inference.}
\label{fig:generative process}
\end{figure}

\section{Dataset Construction}
Existing commonly used social prediction datasets (e.g., BlogCatalog and Flickr \cite{tang2009relational}, YouTube \cite{tang2009scalable}) are designed with a specific task in mind: classifying whether social links exist between pairs of users.
They contain little text or user-attribute information, and are therefore not well suited to evaluate our proposed model. 

Social networks such as Facebook or LinkedIn that support structured profiles would be ideal for the purpose of this work. Unfortunately, they are publicly inaccessible. 
We advert to Twitter. 
One downside of relying on Twitter data is that 
gold-standard information is not immediately available.
Evaluation presented in this paper therefore comes with the flaw that 
 it relies on downstream information extraction algorithms or rule-based heuristics for the attainment of ``gold standards". 
Though not perfect as ``gold-standards" extraction algorithm can be errorful, such a type of evaluation comes with the advantage  that it can be done automatically to compare lots of different systems for development or tuning in relatively large scale.
Meanwhile
the way that our dataset is constructed   
 gives important insights about how the proposed framework can be applied when some structured data is missing\footnote{Facebook and LinkedIn do come with the ideal property of supporting structured user profiles but hardly anyone fills these out.} and how we can address these challenges by directly from unstructured text, making this system applicable to a much wider scenario.

\subsection{User Generated Texts and Graph Networks}
We randomly sample a set of Twitter users, discarding non-English tweets and users with less than 100 tweets.
For each user, we crawl their published tweets and following / follower network
using the publicly available Twitter API.
This results in a dataset of 75 million tweets.
\subsection{User Attributes}
Unlike social networking websites such as Facebook, Google+ and LinkedIn, Twitter does not support structured user profile attributes such as 
gender, education and employer.
We now briefly describe how we enrich our dataset with
user attributes (location, education, gender) and user relations (friend, spouse).
Note that, the goal of this section is to construct a relatively comprehensive dataset for the propose of model evaluation
rather than developing user attribute extraction algorithms.

\subsubsection{Location}\label{sec:location}
We first associate one of the 50 US states with each user.
In this paper, we employ a rule-based approach for user-location identification.\footnote{
While there has been a significant work on geolocation inference,
(e.g., \cite{cheng2010you,conover2013geospatial,davis2011inferring,onnela2011geographic,sadilek2012finding}), the primary goals of this work are to develop
user representations based on heterogenous social signals.  We therefore take a simple high-precision, low-recall approach to identifying user locations.}.
We select all geo-tagged tweets from a specific user, 
and say an location $e$ corresponds to the location of the current user $i$ if it satisfies
the following criteria, designed to ensure high-precision:
(1) user $i$ published more than 10 tweets from location $e$ .
(2) user $i$ published from location $e$ in at least three different months of a year.
We only consider locations within the United States and entities are matched to state names
via Google Geocoding. 
 In the end, we are
able to extract locations for 1.1$\%$ of the users from our dataset.
\subsubsection{Education/Job}
We combine two strategies to harvest gold-standard labels for users' occupation and educational information.
We use The Standard Occupational
Classification (SOC)\footnote{\url{http://www.ons.gov.uk/ons/
guide-method/classifications/
current-standard-classifications/
soc2010/index.html}} to obtain a list of occupations, a approach similar to   \newcite{preoctiuc2015studying,preotiucanalysis}.
\footnote{ SOC is
a UK government system developed by the Office of National Statistics that groups jobs into
23 major categories (for example:  or Engineering Occupations or Legal Occupations), each of which is associated with a set of specific job titles (e.g., mechanical engineer and pediatrist for Professional Occupations).
We construct a lookup table from 
 job occupations to SOC  and apply a rule-based mapping strategy to 
retrieve a users' occupation information based on the free-text user description
field from their Twitter profile. Note that this approach introduces some bias: users with high-profile occupations are more likely to self-disclose their occupations than users with less prestigious occupations.}
A user is assigned an occupation if one of the keywords from the lookup table is identified in his/her profile.
$12\%$ percent of users' occupations are identified using this strategy. 
\subsubsection{Gender}
Following a similar strategy as was described for location and occupation, we take implement a simple high-precision approach for obtaining gold-standard 
user gender information.  We leverage the national Social Security Gender Database
(SSGD)\footnote{\url{http://www.ssa.gov/oact/babynames/names.zip}} to identify users' gender based on their first names.
SSGD contains first-name records annotated for gender for every US birth since
1880 A.D\footnote{Again, we note the large amount of related work on predicting gender of social media users (e.g.,
\cite{burger2011discriminating,ciot2013gender,pennacchiotti2011machine,tang2011s}.)
studying whether high level tweet features (e.g., link, mention, hashtag
frequency) can help in the absence of highly-predictive user name information. As mentioned before, we do not adopted 
machine learning algorithms for attribute extraction.
}.   
Using this database we assign gender to $78\%$ of users in our dataset.

\section{Experiments}
We now turn to experiments on using global inference  
to augment individual local detectors to infer user's attributes,
relations and preferences.
All experiments are based on datasets described in the previous
sections. 
We performed 3 iterations of stochastic gradient descent training over the collected dataset to learn embeddings.
For each task, we separate the dataset into 80\% for training 10\% development and 10\% for testing.

For comparison purposes, neutral models that take into account only part of the training signals presented naturally constitute baselines.
We also implement feature-based SVM models as baselines for the purpose of demonstrating strength of neural models. 
For neural models, we set the latent dimensionality $K$ to $500$.
Pre-trained word vectors are used based on the word2vec package.\footnote{\url{https://code.google.com/p/word2vec/}}  Embeddings are trained on a Twitter dataset consisting of roughly 1 billion tokens.

\subsection{Friend-Relation (Graph Link) Prediction}
Twitter supports two types of following patterns, \textsc{following} and \textsc{followed}. 
We consider two users as friends if they both follow each other.
The friendship relation is extracted straightforwardly from the Twitter network. 
Models and baselines we employ include:
\begin{tightitemize}
\item {\it All}: The proposed model that considers text, graph structure and user attributes.
\item {\it Only Network}: A simplified version
of the proposed model 
 that only used the social graph structure to learn user representations a
 Note that by making this simplification, the model is similar to DeepWalk \cite{perozzi2014deepwalk} with the exception that we adopt negative sampling rather than hierarchical softmax.
\item {\it Network+Attribute}:  Neural models that consider social graph and relation/entity information. 
\item {\it Network+text}:  Neural models that consider social graph and text information. 
\end{tightitemize}

Performance for each model is shown  in Table \ref{friend}. 
As can be seen, taking into account 
 different types of social signals yields progressive performance improvement:
 {\it Graph+Attribute} performs better than {\it only graph}, and {\it All}, which consider all different types of social signals is better than  {\it Graph+Attribute} .
\begin{table}
\centering
\small
\begin{tabular}{cc}
Model&Accuracy\\\hline
All&0.257\\
Only Network&0.179\\
Network+Attribute&0.198\\
Network+Text&0.231\\
\end{tabular}
\caption{Accuracy for different models on friend relationship prediction from social representations.}
\label{friend}
\end{table}

\subsection{User Attributes: Job Occupation}
We present experimental results for job classification based on user-level representations. 
Evaluation is performed on the subset of users whose job labels are identified by the rule-based approach described in the previous section. 
Our models are trained to classify the  top-frequent 10 categories of job occupations

Again, {\it all} denotes the model that utilizes all types of information. 
Baselines we consider include:
\begin{tightitemize}
\item {\it Text-SVM}: We use SVM-Light package to train a unigram classifier that only considers text-level information. 
\item {\it Only Network}: A simplified version of the proposed model that trains user embedding based on network graph and occupation information. 
\item {\it Network+Text}: Embeddings are trained from user-generated texts and network information. 
\end{tightitemize}

Experimental results are illustrated in Table \ref{job}. 
As can be seen, user generated content offers informative evidence about job occupation. We also observe that considering network information  yields significant performance improvement 
 due to the homophily effect, which has been spotted in earlier work \cite{li2014weakly}. 
Again, the best performing model is the one that considers all sorts of evidence. 

\begin{table}
\centering
\small
\begin{tabular}{cc}
Model&Accuracy\\\hline
All&0.402\\
Only Network&0.259\\
SVM-text&0.330\\
Network+Text&0.389\\
\end{tabular}
\caption{Accuracy for different models on 9-class job occupation prediction from social representations.}
\label{job}
\end{table}

\subsection{User Attribute: Gender}
\begin{table}
\centering
\small
\begin{tabular}{cc}
Model&Accuracy\\\hline
All&0.840\\
Only Network&0.575\\
Only Text&0.804\\
SVM-text&0.785\\
Attribute+Text&0.828\\
\end{tabular}
\caption{Accuracy for different models on 9-class job occupation prediction from social representations.}
\label{gender}
\end{table}

We evaluate gender based on a dataset of 10,000 users (half male, half female).
The subset is 
 drawn from the users whose 
gold standard gender labels are assigned  by the social-security  system described in the previous section.
Baselines we employ include: {\it SVM-Text}, in which   a SVM binary classification model is trained  on unigram features; {\it Only-Text}, in which user representations are learned only from texts; {\it Only-Network}, in which user representations are only learned from social graphs; and {\it Text+Relation}, in which representations are learned from text evidence and relation/entity information.

The proposed neural model achieves an accuracy value of  0.840. 
which is very close to the best performance that we are aware of described in \newcite{ciot2013gender}, which achieves 
an accuracy of   0.85-0.86 on a different dataset. However, unlike in \newcite{ciot2013gender}, the proposed model does not require massive efforts  in  feature engineering, 
which involves 
 collecting a wide variety of manual features such as entities mentioned, links, wide range of writing style features, psycho-lingsuitic  features, etc. 
 This demonstrates  
 the flexibility and scalability 
 of deep learning models
 to utilize and integrate different types of social signals 
  on inference tasks over social networks,

User-generated contexts offer significant evidence for gender.  
Again, we observe that leveraging all sorts of social evidence leads to the best performance. 

Experimental results are shown in Figure \ref{gender}. As can be seen, network information does significantly help the task of gender identification, only achieving slightly better performance than random guess. Such an argument is reinforced by the fact that  {\it Text+Relation} yield almost the same performance as model {\it all}, which takes Text+ Relation+ network information. 

\begin{table}
\centering
\small
\begin{tabular}{cc}
Model&Accuracy\\\hline
All&0.152\\
Only Network&0.118\\
Only Text&0.074\\
Network+Text&0.120.\\
Attribute+Text&0.089\\
\end{tabular}
\caption{Accuracy for different models location prediction from social representations.}
\label{location}
\end{table}

\subsection{User Attribute: Location}
The last task we consider is 
 location identification.
 Experiments are conducted 
  on users whose locations have been identified using the rule-based approach described in the previous section. 
The task can be thought of as a 50-class classification problem
and the goal is to pick 
 one from the 50 states (with random-guess accuracy being 0.02). 
We employ baselines similar  to earlier sections: {\it only-text}, 
{\it only network},
{\it text+attribute} and {\it text+network}. 

Results are presented in Table \ref{gender}: both text and network evidence  provide
informative evident  about where a user lives, leading to better performances. 
Again, the best performance is obtained when all types of social signals are jointly considered

\subsection{Examples for Group Behavior Inference}
Given the  trained classifiers  (and additionally trained \textsc{like-dislike} classifiers with details shown in the Appendix), we are able to  perform group behavior inference. 
Due to the lack gold standard labeled dataset, we did not perform  evaluations, but rather list a couple of examples  to give readers a general sense of the proposed paradigm:
\begin{itemize}
\item P(isMale$\Rightarrow$isEngineer)=3.5$\times$ P(isFemale$\Rightarrow$isEngineer)
\item P(isMale,LiveInCalifornia$\Rightarrow$isEngineer)=\\
.~~~~~~~~~~~6.8$\times$$\cdot$ P(isFemale,LiveInCalifornia$\Rightarrow$isEngineer)
\item P(LiveInColorado$\Rightarrow$LikeOmelet)=\\
.~~~~~~~~~~~1.4$\times$P(LiveInCalifornia$\Rightarrow$LikeOmelet)
\item P(LiveInTexas$\Rightarrow$LikeBarbecue)=\\
.~~~~~~~~~~~1.7$\times$P(LiveInCalifornia$\Rightarrow$LikeBarbecue)
\end{itemize}

\section{Related Work}
\label{related_work}
Much work has been devoted to automatic user attribute inference given  social signals.
For example, \cite{rao2010classifying,ciot2013gender,conover2011political,sadilek2012finding,hovy2015user}
focus on how to infer individual user attributes such as age, gender, political polarity, locations, occupation, educational information (e.g., major, year of matriculation) given user-generated contents or network information. 

Taking advantage of large scale user information, recent research has begun exploring logical reasoning 
over the social network (e.g., what's the probability that a New York City resident is a fan of the New York Knicks).
Some work \cite{li2014inferring,wang2013programming} relies on logic reasoning paradigms such as Markov Logic Networks (MLNs) \cite{richardson2006markov}.

Social network inference usually takes advantage of  the fundamental 
propoety of homophily \cite{mcpherson2001birds}, which states that people sharing
similar attributes have a higher chance of becoming friends\footnote{Summarized by the proverb ``birds of a feather
flock together" \cite{al2012homophily}.}, and conversely
friends (or couples, or people living in the same location) tend to share more attributes. 
Such properties have been harnessed for applications like community detection \cite{yang2013community} and user-link prediction \cite{perozzi2014deepwalk,tang2009relational}.

The proposed framework also focuses on attribute inference, which can be reframed as relation identification 
problems, i.e., whether a relation holds between a user and an entity. This work is thus related to 
a great deal of recent researches on
 relation inference (e.g.,
\cite{gu2015traversing,wang2014knowledge,riedel2013relation}). 

Our work is inspired by classic work on spectral 
learning for graphs 
e.g., \cite{kunegis2009learning,estrada2001generalization}   and on
recent research \cite{perozzi2014deepwalk,tang2015line}
that
 learn embedded representations for a graph's vertices.  Our model extends this work
by modeling not only user-user network graphs, but also incorporating 
diverse social signals including unstructured text, user attributes, and relations, 
enabling more sophisticated inferences and offering an integrated model of homophily in social relations.

\section{Conclusions}
We have presented a deep learning framework for
learning social representations, inferring the latent attributes of people online.
Our model offers a new way to jointly integrate noisy heterogeneous cues  from
people's text, social relations, or attributes into a single latent representation.
The representation supports an inference algorithm that can
solve social media inference tasks related to both individual and group behavior,
and can scale to the large datasets necessary to provide practical 
solutions to inferring huge numbers of latent facts about people.

Our model has the ability to incorporate various kinds of information, and it
increases in performance as more sources of evidence are added. 
We demonstrate
 benefits on a range of social media inference tasks, including predicting user gender,
occupation, location and friendship relations.

Our user embeddings naturally capture
the notion of homophily---users who are friends, have similar attributes, 
or write similar text are represented by similar vectors.
These representations could benefit a wide range of downstream applications, such as
friend recommendation, targeted online advertising, and 
further applications in the computational social sciences.  
Due to limited 
publicly 
accessible datasets, we only conduct our experiments on Twitter. However, our algorithms hold potentials to yield more benefits by combining different attributes from online social media, such as Facebook, Twitter, LinkedIn, Flickr\footnote{Images can be similarly represented as vector representations obtained from CovNet \cite{krizhevsky2012imagenet}, which can be immediately incorporated into the proposed framework.},

\bibliographystyle{acl_natbib}
\bibliography{social}  
\noindent  {\Large {\bf Appendix}} \\

{\bf Predicting Preference: Likes or Dislikes}:

Here we describe how we extract user preferences, namely, \textsc{like(usr,entity)} and \textsc{dislike(usr,entity)}. 

Similar 
to a wide range of work on sentiment analysis  
(e.g., \cite{choi2005identifying,kim2006extracting,yang2013joint,agarwal2011sentiment,kouloumpis2011twitter,wangs,pak2010twitter,saif2012semantic}),
our goal is to identify sentiment and extract the target and object 
that express
 the sentiment. 
Manually collecting training data
is problematic because (1) tweets talking about what the user \textsc{likes/dislikes} are very sparsely distributed among
the massive number of topics people discuss on Twitter  and 
(2) tweets expressing what the user \textsc{likes} exist in a great variety of scenarios and forms.

To deal with data sparsity issues, we collect training data by combining {\em semi-supervised information harvesting} techniques \cite{davidov2007fully,kozareva2010learning,kozareva2010not,limajor} and the concept of {\em distant supervision} \cite{craven1999constructing,go2009twitter,mintz2009distant}:

{\bf Semi-supervised information harvesting}: 
We employ a 
 seed-based information-extraction method:
 the model  
 recursively 
uses seed examples to extract patterns, which are then used to 
harvest new examples, which are further used as new seeds to train new patterns.
We begin with pattern seeds including ``I $\_~\_$ like/love/enjoy (entity)", ``I $\_~\_$ hate/dislike (entity)", ``(I think) (entity) is good/ terrific/ cool/ awesome/ fantastic", 
``(I think) (entity) is bad/terrible/awful suck/sucks". 
Entities extracted here should be nouns, which is determined by a Twitter-tuned POS package \cite{owoputi2013improved}.

Based on the harvested examples from each iteration, we train 3 machine learning classifiers: 
\begin{tightitemize}
\item A tweet-level SVM classifier (tweet-model 1) to distinguish between tweets that 
intend to express like/dislike properties and tweets for all other purposes. 
\item A tweet-level SVM classifier (tweet-model 2) to distinguish between like and 
dislike\footnote{We also investigated a 3-class classifier for
like, dislike and not-related, but found the performance constantly underperforms using separate classifiers.}. 
\item A token-level CRF sequence model (entity-model) to identify entities that are
the target of the users like/dislike.
\end{tightitemize}

The SVM classifiers are trained using the SVM$_{light}$ package \cite{joachims1999making} with the following features:
 unigrams, bigrams,  part-of-speech tags and dictionary-derived features based on a subjectivity lexicon \cite{wiebe2005annotating}.

The CRF model \cite{lafferty2001conditional} is trained using the CRF++ package\footnote{\url{https://code.google.com/p/crfpp/}} using the following features:
unigrams, part-of-speech tags, NER tags, capitalization, word shape
and 
 context words within a window of 3 words

The trained SVM and CRF
 models are used to harvest more examples, which are further
used to train updated models.

{\bf Distant Supervision}: The main idea of distant supervision is
to obtain labeled data by drawing on some external sort of evidence. The
evidence may come from a database\footnote{For example, if datasets
says relation \textsc{IsCapital} holds between Britain and London,
then all sentences with mention of ``Britain" and ``London" are
treated as expressing \textsc{IsCapital} relation
\cite{mintz2009distant,ritter2013modeling}.} or common-sense
knowledge\footnote{Tweets with happy emoticons such as :-)  : ) are
of positive sentiment \cite{go2009twitter}.}.  In this work, we
assume that if a relation \textsc{Like(usr, entity)} holds for a
specific user, then many of their published tweets mentioning the
\textsc{entity} also express the \textsc{Like} relationship and are
therefore treated as positive training data.  
Since semi-supervised approaches heavily rely on seed quality 
\cite{kozareva2010not} and the patterns derived by the recursive framework may be strongly influenced by
the starting seeds, adding in examples from distant supervision 
helps increase the diversity of positive training examples.

An overview of the proposed algorithm 
showing how the {\em semi-supervised approach} is combined with {\em distant supervision} is illustrated in Figure \ref{fig1}.

\begin{figure}[ht]
\rule{8cm}{0.03cm}
{\bf Begin}\\
Train tweet classification model (SVM) and entity labeling model (CRF) based on positive/negative data harvested from 
starting seeds.\\
{\bf While stopping condition not satisfied}:
\begin{tightenu}
\item Run classification model and labeling model on raw tweets. 
Add newly harvested positive tweets and entities to the positive dataset.
\item For any user $usr$ and entity $entity$, if relation \textsc{like(usr,entity)} holds,
add all posts published by $usr$ mentioning $entity$ to positive training data. 
\end{tightenu}
{\bf End}\\
\rule{8cm}{0.03cm}
\caption{Algorithm for training data harvesting for extraction user \textsc{like/dislike} preferences.}
\label{fig1}
\end{figure}

{\bf Stopping Condition}: To decide the optimum number of steps
for the algorithm to stop,
we manually labeled a dataset which contains 200 positive tweets (100 like and 100 dislike) with entities.
selected from the original raw tweet dataset rather than the automatically harvested data. 
The dataset contains 800 negative tweets .
For each iteration of data harvesting, we evaluate the performance of the classification models and 
labeling model on this human-labeled dataset, which can be viewed as a development set for parameter tuning. 
Results are reported in Table~\ref{table3}.
As can be seen, the precision score decreases as the algorithm iterates, but the recall rises. 
The best F1 score is obtained at the end of the third round of iteration. 

\begin{table}[h]
\centering
\begin{tabular}{|c|c|l|l|l|}
\hline
\multicolumn{1}{|l|}{}                             & \multicolumn{1}{l|}{} & Pre                       & Rec                       & F1 \\ \hline
\multirow{3}{*}{iteration 1}                       & tweet-model 1         & \multicolumn{1}{c|}{0.86} & \multicolumn{1}{c|}{0.40} &0.55    \\ \cline{2-5} 
                                                   & tweet-model 2         & \multicolumn{1}{c|}{0.87} & \multicolumn{1}{c|}{0.84} & 0.85   \\ \cline{2-5} 
                                                   & entity label          & \multicolumn{1}{c|}{0.83} & \multicolumn{1}{c|}{0.40} &0.54    \\ \hline
\multicolumn{1}{|l|}{\multirow{3}{*}{iteration 2}} & tweet-model 1         & 0.78                      & 0.57                      &0.66    \\ \cline{2-5} 
\multicolumn{1}{|l|}{}                             & tweet-model 2         & 0.83                      &  0.86                         & 0.84   \\ \cline{2-5} 
\multicolumn{1}{|l|}{}                             & entity label          & 0.79                      & 0.60                      & 0.68   \\ \hline
\multirow{3}{*}{iteration 3}                       & tweet-model 1         & 0.76                      & 0.72                      &0.74    \\ \cline{2-5} 
                                                   & tweet-model 2         & 0.87                      & 0.86                      & 0.86   \\ \cline{2-5} 
                                                   & entity label          & 0.77                      & 0.72                      &0.74    \\ \hline
\multirow{3}{*}{iteration 4}                       & tweet-model 1         & 0.72                      & 0.74                      &0.73    \\ \cline{2-5} 
                                                   & tweet-model 2         & 0.82                      & 0.82                      & 0.82   \\ \cline{2-5} 
                                                   & entity label          & 0.74                      & 0.70                    & 0.72   \\ \hline
\end{tabular}
\caption{Performance on the manually-labeled devset at different iterations of data harvesting.}
\label{table3}
\end{table}

For evaluation, data harvesting without distant supervision (denoted by \textsc{no-distant}) naturally constitutes a baseline. Another baseline (denoted by {\it one-step-crf})
 trains a one-step CRF model, which directly decides whether a specific token corresponds to a \textsc{like/dislike} entity rather than making tweet-level decision first. 
Both (\textsc{no-distant}) and \textsc{one-step-crf} rely on the recursive framework and tune the number of iterations on the aforementioned gold standards. 
Test set
consists of 
 100 like/dislike property related tweets (50 like and 50 dislike) with entity labels, which are then matched with 400 negative tweets.
The last baseline we employ is a rule-based extraction approach  using the seed patterns.
We report the best performance model on the end-to-end entity extraction precision and recall.

\begin{table}[h]
\centering
\begin{tabular}{|c|c|c|c|}\hline
Model&Pre&Rec&F1\\\hline
semi+distant&0.73&0.64&0.682\\\hline
no-distant&0.70&0.65&0.674\\\hline
one-step (CRF)&0.67&0.64&0.655\\\hline
rule&0.80&0.30&0.436\\\hline
\end{tabular}
\caption{Performances of different models on extraction of user preferences (like/dislike) toward entities.}
\label{tab2}
\end{table}

From Table \ref{tab2}, we observe performance improvements introduced by  combining user-entity information with distant supervision. 
Modeling tweet-level and entity-level information  yields better performance than moldeing them in a unified model (\textsc{one-step-crf}). 

We apply the model trained in this subsection to our tweet corpora. 
We filter out entities that appear less than 20 times, resulting in roughly 40,000 different entities\footnote{Consecutive entities with same type of NER labels are merged.}.

\end{document}